\documentstyle[prc,preprint,tighten,aps]{revtex}
\newcommand{\beq}{\begin{equation}}
\newcommand{\eq}{\end{equation}}
\newcommand{\beqa}{\begin{eqnarray}}
\newcommand{\eqa}{\end{eqnarray}}

\newcommand{\grvect}[1]{\mbox{\boldmath$#1$}}
\begin {document}
\draft
\title{Microscopic multicluster description of the neutron-rich
helium isotopes}
\author{K. Varga \\
Institute of Nuclear Research of the Hungarian Academy of Sciences,
\\
Debrecen, P. O. Box 51, H--4001, Hungary
\\
and
\\
Y. Suzuki and Y. Ohbayasi
\\
Physics Department, Niigata University, Niigata 950--21, Japan}
\date{\today}
\maketitle
\begin{abstract}
\noindent
The neutron halo structure of $^{6}$He and of $^{8}$He is studied
in a microscopic three-body and five-body model, respectively.
Various cluster arrangements are included
to embody a variety of correlations between the clusters. The
intercluster wave function is determined with the stochastic variational
method.  The $^{6}$He and $^{8}$He energies are reproduced with the same
effective force very well.  The radii and densities are compared with the
results of empirical analyses.  The calculated difference of the
neutron and proton radii exceeds 0.8 fm, confirming the thick neutron
skin.  The two-neutron removal spectroscopic amplitudes are calculated
and used for evaluating the momentum distribution and the $\beta$
decay spectrum, which are sensitive to the halo structure.  A fair
agreement with experiment is obtained.
\end{abstract}


\narrowtext

\section{Introduction}

The measurement of the interaction cross section \cite{intcross} and
the momentum distribution of fragments in neutron removal reactions
\cite{momdist}
of the light unstable neutron-rich nuclei revealed the unusually
large matter radii, and led to
the concept of the so called halo \cite{halo} or neutron skin \cite{skin}
structure of such nuclei.
The description of the halo nuclei has since become a challenge for
theoretical nuclear model calculations.
\par\indent
In a previous paper \cite{previous}, we studied the ground states
of the nuclei $^6$He and $^8$He.
The $^8$He nucleus, with its extremely large neutron excess, is one
of the most intriguing examples of the halo nuclei.
To give a realistic description of this nucleus, one has to go
beyond the $^6$He+n+n-type or $^4$He+(2n)+(2n)-type three-body
models, because in the former the assumption of the $^6$He inert core is
obviously questionable, while in the latter the assumption of the
dineutron clusters may exaggerate the neutron correlation.
In fact, it has recently been shown \cite{skoyb} that a picture for $^8$He
consisting of $^6$He plus a two-neutron halo is not justified by the
fragmentation cross section data of the helium isotopes.
\par\indent
In our microscopic multicluster approach  the helium isotopes
comprise an $\alpha$ cluster and single-neutron ``clusters''.
In this model, the intercluster wave function is a superposition
of terms of different relative coordinate arrangements, and
each term is a product of nodeless harmonic oscillator functions with
different width parameters.
The flexibility of the model enables us to include various kinds
of important correlations in the halo nuclei on an equal footing.
To determine the combination coefficients,
the ``stochastic variational method'' was used,
which is an extraordinarily
good procedure to select the important states.
\par\indent
The present model has some advantages over the previous approaches.
Unlike the cluster-orbital shell model \cite{COSM} and the three-body
\cite{threebody,Zhukov} approaches,  we do not have to use the
passive-core-plus-outer-nucleons approximation and we take into
account the Pauli principle properly. Although the halo nuclei
have been investigated by different versions of the microscopic cluster
model \cite{clmodel}, our approach is unique as it treats more
than three clusters microscopically.
\par\indent
The aim of this paper is to demonstrate that this model provides
realistic wave function  of halo nuclei, that is, to show that the
physical properties extracted from the wave functions are close
to the experimental results.
\par\indent
The plan of this paper is as follows. In sect. 2 we briefly sketch
the model used. In sect. 3 we present the results of the numerical
calculations. Sect. 4 contains a summary and some conclusions.

\section{The microscopic multicluster model}
In this section we briefly outline the most important part of
our formalism. A more detailed description is given in our previous paper.
To describe the system of $\alpha$ plus $n$ single neutron clusters, we
build up a trial function which is a sum over various cluster arrangements
$\mu$, each associated with a particular set of intercluster Jacobi
coordinates ${\grvect \rho}_1^\mu,...,{\grvect \rho}_n^\mu$.
The spins of the clusters are coupled to $S$, and the orbital
angular momenta $l_i\equiv l_i^\mu$ belonging to the Jacobi coordinates
${\grvect \rho}_i^\mu$
are coupled to L. The wave function of the intercluster motion
is approximated by a linear combination of nodeless harmonic oscillator
functions of different size parameters:
\beq
\Gamma^{\mu}_{ikl_i m_i}(\grvect{\rho}^{\mu}_i)=
\left[{2^{2l_i+7/2}(\nu^{\mu}_{ik})^{l_i+3/2}\over\sqrt{\pi}(2l_i+1)!!}
\right]^{1/2}
(\rho^{\mu}_i)^{l_i} {\rm exp}[-\nu^{\mu}_{ik}(\rho^{\mu}_i)^2]
Y_{l_i m_i}(\hat{\grvect{\rho}}^{\mu}_i),
\label{Gamma}
\eq
where $\nu^{\mu}_{ik}$ is the $k$th size parameter
of the $i$th relative motion in the cluster arrangement $\mu$.
\par\indent
The wave function belonging to arrangement $\mu$ and angular
momenta $[S,(l_1...l_n)L]JM$ can be written as
\beq
\Psi ^{\mu}_{[S,(l_1...l_n)L]JM}
=\sum_{K} C^{\mu}_{K,l_1,...,l_n} {\cal A}\left \{\left [\Phi _S
{\Gamma }^{\mu}_{K(l_1...l_n)L}
(\grvect{\rho}^{\mu}_1,...,\grvect{\rho}^{\mu}_n)
\right ]_{JM}\right \},
\label{trialfnterm}
\eq
where ${\cal A}$ is the intercluster antisymmetrizer,
$\Phi _{SM_S}$ is a vector-coupled product of the  intrinsic wave
function of the $\alpha$ particle,
$\Psi _{^4{\rm He}}$, and the $n$ neutron spin-isospin functions
$\Phi^i$. The function ${\Gamma}^{\mu}_{K(l_1...l_n)L}$ is a vector-coupled
product of the intercluster relative functions
$\Gamma^{\mu}_{ikl_i m_i}(\grvect{\rho}^{\mu}_i)$, where $K$ stands for
the set of the indices $\{k_1,...,k_n\}$ of the size parameters.
The sequence of angular momentum coupling is chosen so as to follow
the pattern of the Jacobi coordinates.
The intrinsic wave function, $\Psi_{^4{\rm He}}$, is constructed from a
harmonic oscillator Slater determinant with size parameter $\nu$.
\par\indent
The variational trial function
$\Psi_{^6{\rm He}}$ ($\Psi_{^8{\rm He}}$) of $^6$He ($^8$He)
is a combination of different arrangements and
the intercluster angular momenta:
\beq
\Psi=\sum_{\mu}\sum_{l_1...l_n}\Psi^{\mu}_{[S,(l_1...l_n)L]JM}.
\label{trialfn}
\eq
This function contains a great number of terms, due not only
to the different arrangements and angular momenta but, especially,
to the various size parameters.
Owing to this fact our trial function becomes so flexible that it can
describe both various types of correlation between the clusters
and the spatially extended halo structure.
In the previous paper we tested various methods to select
$\nu_{ik}^\mu$ that span most adequately the state
space, while the dimension of the basis is kept feasible.
The most efficient procedure found is the following. We generate
size parameter sets randomly chosen from an interval which
corresponds to the physically important region. The parameter sets
that satisfy an admittance condition are selected to be
basis states. We admitted a candidate if it, together with the
previously selected basis states, lowers the energy more than a preset
value, $\epsilon$=0.005 MeV. To avoid declaring convergence prematurely
during this procedure, after 10 failed attempts, we removed the condition,
and accepted the next candidate. The repetition of this procedure
yields excellent numerical convergence in energy and reduces the number
of trial terms considerably.
\par\indent
In this paper we use the same potential as in the previous one, i.e.
the central effective interaction of
Thompson, LeMere and Tang \cite{TLT}, restrict ourselves to
$L=0, S=0$ and neglect the Coulomb force. This interaction is
chosen so as
to reproduce the most important low-energy nucleon-nucleon scattering data and
therefore it does not bind the dineutron.
The parameter $u$ of the interaction was set to reproduce
qualitatively the $p_{3/2}$
$\alpha$-nucleon phase shifts and the binding energy of $^6$He
$(u=1.15)$. The nucleus $^7$He does not exist in nature,
and this interaction does not produce it in a bound state.
\par\indent
In our previous study the size parameter of the $\alpha$ particle was chosen
so as to minimize the energy of the $\alpha$ particle
($\nu=0.303$ fm$^{-2}$). This
choice, however, gives smaller radii than the empirical values by about
$8\%$ for the $\alpha$ particle and for $^8$He.
In the present work we adjusted the radius of the $\alpha$ particle
to its realistic value ($\nu=0.270$ fm$^{-2}$), allowing the
$\alpha$ particle to be slightly less bound and expecting the
radius of $^8$He to come closer to the experimental findings
(the experimental data
and our results are compared in Table 1).
After fixing the parameter $u$ and $\nu$  the model does not
contain any free parameter for the description of $^8$He.

\section{Results}

\par\indent
In the calculation for $^6$He we used the same configurations
as in the previous work, that is, $l_1=l_2=0$ partial waves for the
$\alpha+(nn)$ configuration $(T_{00})$ (the T$_{11}$ channel is
Pauli-forbidden) and
$l_1=l_2=0,1$  partial waves for $(\alpha n)n$ $(Y_{00},Y_{11})$.
To elucidate the role of the inclusion of the different cluster arrangements,
we repeated our calculation using only the $Y$-type channels but including
$(Y_{22})$ and $(Y_{33})$ as well. The energy falls as -0.382 MeV
(in $\{Y_{00}\}$), -0.601 MeV (in $\{Y_{00},Y_{11}\}$),
-0.823 MeV (in $\{Y_{00},Y_{11},Y_{22}\}$) and -0.923 MeV
(in $\{Y_{00},Y_{11},Y_{22},Y_{33}\}$). At the same time, the energy
in the $\{T_{00},Y_{00},Y_{11}\}$ model space is -0.990 MeV, and
inclusion of higher partial  waves ($\{T_{00},T_{22},Y_{00},
Y_{11},Y_{22},Y_{33}\}$)  changes it by less than 5 keV (-0.994 MeV).
We can conclude, in accordance with Ref. \cite{Kamimura}, that
with the inclusion of different rearrangement terms the higher partial
waves may safely be omitted.
\par\indent
For the description of $^8$He we selected
the arrangements found dominant previously, but complemented
them  by some $l=1$ partial waves as well. Thus the following
six channels were coupled:
\par\noindent
$\{[(\alpha n)n]n\}n$, $[(\alpha n)n](nn)$, $[\alpha (nn)](nn)$,
 $\{[\alpha (nn)]n\}n$, $\{[(\alpha n)_1n]_1n\}n$,
$[(\alpha n)_1n]_1(nn)$
\par\noindent
The lower indices indicates that $l=1$ partial waves were
taken in the corresponding relative motion.
These partial-wave pairs were then coupled to total
angular momentum $L=0$. These configurations of $^8$He, in fact,
contain the most important configurations of $^6$He.
Using the procedure
described above the energy minimum
has been reached after 124, 91, 57, 64, 31, 19 steps
in the successive channels, lowering the energy, in turn, as
$-2.49$, $-3.07$, $-3.23$, $-3.28$, $-3.29$, $-3.30$ MeV. By
consulting Table 1,
we see that $^8$He is overbound by about 200 keV. This little overbinding
is partly due to the fact that the interaction chosen overbinds
$^6$He by 20 keV. If it were fine tuned to reproduce the energy of the
$^6$He, a more perfect agreement would be obtained, as discussed in Ref.
\cite{previous}.
\par\indent
We have calculated the rms point matter radii of $^6$He and $^8$He.
The results are shown in Table 1.
Effort was made to extract the radii of these nuclei by using the
relevant interaction cross section data. However, the extracted values
diagree, as listed in Table 1, according to the type of model analysis
performed. (See Ref. \cite{OYS}  for a  detailed theory of
evaluating the various fragmentation cross sections for the halo nuclei
in the framework of the Glauber theory \cite{Glau}).
The empirical estimates for the difference of the
proton and neutron rms radii (thickness of the neutron skin)
are 0.87 fm for $^6$He and 0.93 fm for $^8$He (by Ref. \cite{skin}) and
0.4 fm for $^6$He and 0.49 fm for $^8$He (by Ref. \cite{fr}).
The calculated differences of the proton and neutron radii
(0.87 fm for $^6$He and 0.82 fm for $^8$He) are
close to the recent results of Ref. \cite{skin}, showing the thick
neutron skin.
The only striking discrepancy between the empirical estimates and our model
is that our calculation gives smaller matter radius for $^8$He than  for
$^6$He. We shall discuss this point in section 4.
To illustrate that the model space has been adequately spanned,
we note here that the radius of $^8$He, after about first 150 steps,
did not change significantly.
\par\indent
The point nucleon density distribution has
also been determined. Although there is no experimental density
distribution available, we can compare our results to the
qualitative empirical estimates.
In a recent paper \cite{skin}  Tanihata et. al. have
attempted to deduce  the point nucleon
density distributions of $^6$He and $^8$He.
They assumed a harmonic oscillator density distribution combining
the $0s$ and the $0p$ orbitals with different harmonic oscillator
size parameters. They calculated the interaction cross
sections of these nuclei using the optical limit of the Glauber model.
The oscillator size parameters were determined by fitting the
results to the experiments. However, referring to Ref.
\cite{OYS}, we note that the optical limit approximation is not very accurate
in evaluating the cross sections of the halo nuclei characterized by
spatially extended density distributions.
We present our density distributions
and their  results in Fig 1.a and 1.b. The calculated density
distributions are not far from the empirical ones and reproduce the most
pronounced property, namely the distribution of neutrons extending
far beyond that of protons.
\par\indent
Another important quantity that helps to reveal information about the wave
function of the halo nuclei is
the two-neutron removal spectroscopic amplitude
of $^6$He and $^8$He defined by (see e.g. \cite{VargaLovas})
\beq
g_{^6{\rm He}}({\bf r},{\bf R})=
\left({6!\over 4! 1! 1!} \right)^{1/2}
\langle
\Psi_{^4{\rm He}} \delta({\grvect \rho}_1-{\bf r})
\delta({\grvect \rho}_2-{\bf R}) \vert
\Psi_{^6{\rm He}} \rangle
\eq
and
\beq
g_{^8{\rm He}}({\bf r},{\bf R})=
\left({8!\over 6! 1! 1!} \right)^{1/2}
\langle
\Psi_{^6{\rm He}} \delta({\grvect \rho}_1-{\bf r})
\delta({\grvect \rho}_2-{\bf R}) \vert
\Psi_{^8{\rm He}} \rangle.
\eq
Here $\grvect \rho_1$ is a Jacobi coordinate connecting two
neutrons and $\grvect \rho_2$ is the Jacobi coordinate connecting
the c.m. of these two neutrons with the c.m. of the rest;
$\bf r$ and $\bf R$ are respective parameter coordinates.
Note that the intrinsic wave function of $^6$He , $\Psi_{^6{\rm He}}$, in
eqs. (4) and (5) is the same. The norm squares of these amplitudes
(the two-neutron removal spectroscopic factors) are
$\langle g_{^6{\rm He}} \vert g_{^6{\rm He}} \rangle=1.39$ and
$\langle g_{^8{\rm He}} \vert g_{^8{\rm He}} \rangle=1.36$.
These spectroscopic factors are larger than unity
but this is not surprising as
the two-neutron spectroscopic factors in the pure harmonic oscillator
limit, i.e. when the wave function of the relative motions are
shell model harmonic oscillator functions of the same size parameters,
are ${13 \over 8}$ and ${ 13 \times 25 \over 8 \times 24 }$
for $^6$He and $^8$He, respectively \cite{VargaLovas}.
\par\indent
The radial spectroscopic amplitudes
\beq
g_{^6{\rm He}}^0(r,R)=
\int d {\hat {\bf r}}
Y_{00}({\hat {\bf  r}})^{\ast}
\int d{\hat {\bf R}}
Y_{00}({\hat {\bf R}})^{\ast}
g_{^6{\rm He}}({\bf r},{\bf R})
\eq
and
\beq
g_{^8{\rm He}}^0(r,R)=
\int d{\hat {\bf r}}
Y_{00}({\hat {\bf r}})^{\ast}
\int d{\hat {\bf R}}
Y_{00}({\hat {\bf R}})^{\ast}
g_{^8{\rm He}}({\bf r},{\bf R})
\eq
are plotted in Figs. 2a and 2b, respectively.
We note here that the norm of these amplitudes are
$\langle g_{^6{\rm He}}^0 \vert g_{^6{\rm He}}^0 \rangle=1.31$ and
$\langle g_{^8{\rm He}}^0 \vert g_{^8{\rm He}}^0 \rangle=1.32$, thus
about 95\% of the spectroscopic amplitudes lie in the subspace
of the $l_1=0, l_2=0$ partial waves.
Our model gives qualitatively the same spectroscopic amplitudes for $^6$He
as the three-body models \cite{Zhukov,clmodel}, that is, it shows
two prominent peaks: the ``di-neutron'' peak ($r=1$ fm, $R=4$ fm)
where the two neutrons
are close to each other, and the ``cigar-like'' configuration, with two
neutrons at almost the opposite sides of the $\alpha$ particle
($r=2.75$ fm, $R=2$ fm).
The two-neutron removal spectroscopic amplitude of $^8$He has a
shape rather
similar to that of $^6$He, although it is spatially
more compact. The positions of its peaks are almost the same
as in the case of $^6$He, the magnitudes of the peak are, however,
about 1.25 times higher than those of $^6$He.
\par\indent
The two-neutron spectroscopic amplitude defined above can be tested by
experimental data such as the momentum distribution of the fragment in the
two-neutron removal  reactions and the $\beta$-decay spectrum.
\par\indent
The normalized momentum distribution of the fragment
in one direction is given by
\beq
\rho(k_x)=
{1 \over \langle g \vert g \rangle}
\int\int d k_y d k_z d{\bf q}
\vert {\hat g}({\bf q,\bf k})\vert^2,
\eq
where
${\hat g}({\bf q,\bf k})$ is defined as
\beq
{\hat g}({\bf q,\bf k})=\int\int {d{\bf r} d{\bf R}
\over (2\pi)^3} e^{-i{\bf q} {\bf r}-i{\bf k} {\bf R}}
g({\bf r},{\bf R}).
\eq
In this equation we assumed that the momentum distribution measured
in fragmentation reactions is proportional to the momentum
distribution $\vert {\hat g}({\bf q,\bf k})\vert^2$
of the fragment and the neutrons in the projectile.
In general, the reaction dynamics of the two-neutron removal process
gives modifications to eq. (8).
The change for the longitudinal momentum distribution
was found to be small at high energies \cite{OSY}.
The transverse momentum distribution at high energies has not so far
been analysed fully with the  reaction dynamics included,
but the shape of the transverse and the longitudinal momentum distribution
in the region of low momentum transfer was found to be quite similar
\cite{Fujita}.
Therefore we may assume that the
observed transverse distribution can be safely compared to the momentum
distribution calculated with the spectroscopic amplitude.
The experimental and calculated momentum distributions of the $^4$He fragment
in the two-neutron removal reaction
($^6$He,$^4$He) \cite{momdist} and of the $^6$He fragment in ($^8$He,$^6$He)
\cite{Z} are compared in Figs. 3a and b.
The characteristic features of the momentum distributions are well
reproduced particularly in the case of $^6$He (Fig. 3.a).
The width of the momentum
distribution of $^8$He is slightly wider than in the experiment (Fig. 3.b).
This indicates that the relative motion between $^6$He and the two
neutrons in the halo should  extend spatially a little further.  This
point may be related to the fact that the calculated $^8$He
energy is overbound by about 200 keV.
\par\indent
The two-neutron removal spectroscopic amplitude of $^6$He
can also be used to calculate the $\beta$-decay rate into $\alpha$
and deuteron.
Recent theoretical analyses
\cite{clmodel,CB} have revealed an extreme sensitivity to the
halo description up to distances as large as 15 fm. We show the
$\beta$-decay spectrum obtained with our wave function in order to see
how good it is at large distances.
The $\beta$ delayed deuteron emission transition probability per time
and energy units,$dW \over dE$, can be expressed as
\beq
{dW \over dE}={mc^2 \over \pi^4 v \hbar^2} G_{\beta}^2
f(Q-E)B_{GT}(E),
\eq
where $m$ is the electron mass, $v$ is the relative
velocity between the $\alpha$ particle and the deuteron
and $G_\beta=2.996\times 10^{-12}$ is the
dimensionless $\beta$-decay constant.
The phase space factor, or Fermi integral,
$f$ depends on the kinetic energy $Q-E$ available
for the electron and the antineutrino. The mass
difference between the initial and final particles
is $2.03$ MeV.
The Gamow-Teller reduced transition probability is to a fair
approximation
\beq
B_{GT}(E)=6\lambda^2 \langle F_d(r) \chi_E(R) \vert
g_{^6{\rm He}}^0(r,R) \rangle^2,
\eq
where $F_d(r)$ is the radial part of the deuteron
wave function and $\chi_E$ is the $\alpha$-deuteron
relative wave function in the final state.
See Ref. \cite{clmodel} for details. Fig. 4 displays the effective
function, $R \int dr r^2 F_d(r) g_{^6{\rm He}}^0(r,R) $, as a function
of R.
The asymptotic behavior at large distances is described well enough to
calculate the $\beta$-decay spectrum. The calculated
probability, $dW \over dE$, shown in Fig. 5, reproduces
the experiment \cite{betanew}
reasonably well. The quality of agreement is very similar to
that of Ref. \cite{CB}, where a more sophisticated description of the final
states is used.
\par\indent
The beta delayed triton and $\alpha$ spectra of $^{8}$He have also been
measured \cite{betanew}.  In this case the possibility of the direct decay
mechanism through the $^{5}$He resonance, that is, $^{8}$He $\rightarrow$
t+$^{5}$He$(3/2^{-})$+e$^{-}$+${\bar{\nu_e}}$ and $^{5}$He $\rightarrow$
$\alpha$+n, is ruled
out because a simple symmetry argument predicts vanishing GT matrix elements
for the $\beta$-decay.  A more careful consideration on the $\beta$-decay
mechanism is needed to test the $^{8}$He wave function with use of the
$\beta$-decay spectra of $^{8}$He.

\section{Summary}
The use of the stochastic variational method allows us to treat
the light neutron-rich nuclei as a multicluster system.
This method helps to keep the dimension of the variational
basis low and ensures that the wave function is good enough to describe
the extended neutron halo. In this paper we have calculated
the ground-state energy, wave function and some physical properties
of the nuclei $^6$He and $^8$He in a fully microscopic framework.
The inclusion of various rearrangenment cluster configurations
was shown to be quite effective to describe the halo structure
of the helium isotopes.
\par\indent
We set the only parameter of our simple central nucleon-nucleon
interaction to give the nearly correct ground-state energy
of $^6$He and by using this interaction we reproduced the ground-state
energy of $^8$He  with fair accuracy.
\par\indent
We have calculated the proton, the neutron and the point matter rms
radii of these nuclei. The difference between the neutron
and proton radii in our model, i.e. the thickness of the neutron skin,
is in perfect agreement with the empirical findings.
Unlike empirical estimates, our model gives a little larger rms radius
for  $^6$He than  for $^8$He. This discrepancy may reflects
the effect of the overbinding ($\sim$ 200 keV) of  $^8$He
or the use of a not fully adequate effective interaction (e.g. the
neglect of the spin-orbit force) but, at the same time,
one may be inclined to accept this result since the binding energy
of $^8$He (3.112 MeV) is much larger than the binding energy
of $^6$He (0.975 MeV).
To estimate the effect of the overbinding of $^8$He on the
rms matter radii of $^6$He and $^8$He, we set the parameter
$u$ to $1.145$. This choice gives $0.87$ MeV for the binding energy
of $^6$He, and results less overbinding of $^8$He ($\sim 160$ keV).
While in this calculation the rms matter radius of $^6$He increased
($r=2.48$ fm), that of $^8$He hardly changed, in fact it
decreased by 0.004 fm. Therefore, the overbinding does not seem to be
responsible for the result that the rms matter radius of $^6$He is calculated
to be larger than that of $^8$He.
The calculated result that both the proton and neutron radii
of $^6$He are larger than those of $^8$He may be understood if
one assumes that the distance between the center of mass of the $\alpha$
particle in $^6$He and the center of mass of $^6$He is larger than the
corresponding quantity in the case of $^8$He.
To understand the reason of the discrepancy of the radii, one has to use a
more sophisticated effective interaction, or the extraction of the
proton ore neutron radii from the interaction cross section data
has to be carefully reinvestigated.
\par\indent
We have determined the proton, the neutron and the matter distributions.
The results are in accord with the empirical estimates and the
relativistic mean field calculations \cite{skin}.
\par\indent
The two-neutron removal spectroscopic amplitudes have been calculated  and
used to calculate the momentum distributions of the fragments arising
from the two-neutron removal reaction and the $\beta$-decay of $^6$He into
$\alpha$ and deuteron. The results are in fair agreement with experiment.
\par\indent
\vskip 12pt
This work was supported by the OTKA grants No. 3010 and F4348 (Hungary) and
by a Grant-in-Aid for Scientific Research (No. 05243102) of the Ministry of
Education, Science and Culture (Japan).

\newpage
\begin{table}
\caption{Ground state energies (in MeV),
root mean square matter, proton and neutron
radii (in fm) of $^4$He,$^6$He and $^8$He.
Empirical values of
the radii are taken from Ref. [4]. Values in parenthesis
are from Ref. [15].}
\begin{tabular}[t]{crrcccccc}
\hline\\
&$E_{exp}$&$E_{theor}$&$r^{matter}_{emp}$&$r^{matter}_{theor}$&
$r^{neutron}_{emp}$&
$r^{neutron}_{theor}$&$r^{proton}_{emp}$&$r^{proton}_{theor}$ \\
\hline\\
$^4$He&$-28.30$ &$-$25.243 &1.63 (1.57) &1.63&1.64 (1.57)&1.63&1.64 (1.57)&1.63
 \\
$^6$He&$-$0.975 &$-$0.994  &2.33 (2.48) &2.46&2.59 (2.61)&2.67&1.72 (2.21)&1.80
 \\
$^8$He&$-$3.112 &$-$3.301  &2.49 (2.52) &2.40&2.69 (2.64)&2.53&1.76 (2.15)&1.71
 \\
\hline\\
\end{tabular}
\end{table}
\begin{figure}[h]
\caption{
Empirical (short dashed line) and theoretical (solid line)
proton and neutron density of $^6$He (Fig. 1a) and $^8$He (Fig. 1b).
The long dashed line shows the calculated matter density. The
distributions are normalized to unity.}

\caption{Two-neutron removal spectroscopic amplitudes:
$r R g_{^6{\rm He}}^0(r,R)$ and $r R g_{^8{\rm He}}^0(r,R)$.
The value of the magnitude (in fm) as function of $r$ and $R$ is written
on the contour lines.}

\caption{Experimental (solid line) and theoretical (dashed line)
normalized transverse momentum distribution of the $^4$He
fragment in the ($^6$He,$^4$He) [2] and the $^6$He fragment in the
($^8$He,$^6$He) [19]  two-neutron removal reaction on a carbon target.
The two-neutron removal reaction cross sections are $189\pm14$mb
and $202\pm17$mb for $^6$He and $^8$He, respectively.}

\caption{Absolute value of R \int dr r^2 F_d(r) g_{^6{\rm He}}^0(r,R)$
as a function of R. \ \ \ \ \ \ \ \ \ \ \ \ \ \}

\caption{Transition probability $dW/dE$ per time and energy units
(in s$^{-1}$ MeV$^{-1}$) in the c.m. frame
as a function of the c.m.  energy $E$ (MeV). The experimental
points are from Ref. [21].}

\end{figure}


\begin{thebibliography}{99}
\bibitem{intcross}
I. Tanihata, H. Hamagaki, O. Hashimoto, Y. Shida, N. Yoshikawa, K. Sugimoto,
O. Yamakawa, T. Kobayashi and N. Takahashi, Phys. Rev. Lett. {\bf 55} (1985)
2676
\bibitem{momdist} K. Kobayashi, Nucl. Phys {\bf A538} (1992) 343c
\bibitem{halo}
P. G. Hansen and B. Jonson, Europhys. Lett. {\bf 4} (1987) 409
\bibitem{skin} I. Tanihata, D. Hirata, T. Kobayashi, S. Shimoura, K.
Sugimoto and H. Toki, Phys. Lett. {\bf B289} (1992) 261
\bibitem{previous} K. Varga, Y. Suzuki and R. G. Lovas Nucl.
Phys. A, in press
\bibitem{skoyb} Y. Suzuki, T. Kido, Y. Ogawa, K. Yabana and D. Baye,
Nucl. Phys. {\bf A567} (1994) 957
\bibitem{COSM} Y. Suzuki and K. Ikeda, Phys. Rev. {\bf C38} (1988)
410;\\
Y. Suzuki and J. J. Wang, Phys. Rev. {\bf C41} (1990) 736
\\
Y. Suzuki, Nucl. Phys. {\bf A538} (1991) 395
\bibitem{threebody} V. I. Kukulin, V. M. Krasnopol'sky, V. T. Voronchev and
P. B. Sazonov, Nucl. Phys. {\bf A417} (1984) 128; {\bf A453}
(1986) 365
\\
A. Ghovanlou and D. R. Lehman, Phys. Rev. {\bf C9} (1974) 1730
\bibitem{Zhukov}
M. V. Zhukov, B. V. Danilin, D. V. Fedorov, J. M. Bang, I.
J. Thompson and J. S. Vaagen, Phys. Reports {\bf 231} (1993) 150;
\\
M. V. Zhukov, A. A. Korsheninnikov and M. H. Smedberg, submitted
to Phys. Lett. {\bf B}
\bibitem{clmodel} D. Baye, Y. Suzuki and P. Descouvmont, Prog. Theor. Phys.
{\bf 91} (1994) 271
\bibitem{TLT} D. R. Thompson, M. LeMere and Y. C. Tang, Nucl. Phys. {\bf A286}
(1977) 53
\bibitem{Kamimura} M. Kamimura, Phys. Rev. {\bf A38} (1988) 621
\bibitem{OYS} Y. Ogawa, K. Yabana and Y. Suzuki, Nucl. Phys. {\bf A543}
(1992) 722
\bibitem{Glau} R. J. Glauber, Lectures in theoretiacal physics. Vol. 1
(Interscience, New York, 1959) p. 315
\bibitem{fr} I. Tanihata, T. Kobayashi, O. Yamakawa, S. Shimoura, K. Ekuni,
K. Sugimoto, N. Takahashi, T. Shimoda and H. Sato, Phys. Lett. {\bf B206}
(1988) 592
\bibitem{VargaLovas} K. Varga and R. G. Lovas, Phys. Rev. {\bf C43} (1991) 1201
\bibitem{OSY} Y. Ogawa, Y. Suzuki and K. Yabana, Nucl. Phys {\bf A}, in press
\bibitem{Fujita} T. Fujita and J. H\"ufner, Nucl. Phys. {\bf A343} (1980) 493
\bibitem{Z} T. Kobayashi, O. Yamakawa, K.Omata, K. Sugimoto, T. Shimoda, N.
Takahashi and I. Tanihata, Phys. Rev. Lett. {\bf 60} (1988) 2599
\bibitem{CB} A. Cs\'ot\'o and D. Baye, Phys. Rev. {\bf C} (1994)
\bibitem{betanew} M. J. G. Borge
L. Johannsen, B. Jonson, W. Kurcewicz, T. Nilsson, G. Nyman,K. Riisager,
O. Tengblad, and K. Wilhelmsen, Nucl. Phys. {\bf A560} (1993) 664
\end{thebibliography}
\end{document}